\documentclass{iau}
\usepackage{graphicx}
\usepackage{amsmath}

\title[Pebble Delivery for Inside-Out Planet Formation] %% give here short title %%
{Pebble Delivery for\\Inside-Out Planet Formation}

\author[Xiao Hu, Jonathan C.\ Tan, \& Sourav Chatterjee]   %% give here short author list %%
{Xiao Hu$^1$ Jonathan C. Tan$^1$
 \and Sourav Chatterjee$^2$}

\affiliation{$^1$Department of Astronomy, University of Florida, 32611, Gainesville, Florida, USA. \\ email: {\tt ustcxhu@ufl.edu,  \tt jt@astro.ufl.edu} \\[\affilskip]
$^2$Center for Interdisciplinary Exploration and Research in Astrophysics (CIERA)\\
Department of Physics \& Astronomy, Northwestern University, Evanston, IL 60208, USA. \\email: {\tt sourav.chatterjee@northwestern.edu}}

\pubyear{2014}
\volume{310}  %% insert here IAU Symposium No.
\pagerange{119--126}
\jname{Complex Planetary Systems}
\editors{xxxx.}
\begin{document}

\maketitle

\begin{abstract}
Inside-Out Planet Formation (IOPF; Chatterjee \& Tan 2014, hereafter
CT14) is a scenario for sequential {\it in situ} planet formation at
the pressure traps of retreating dead zone inner boundaries (DZIBs)
motivated to explain the many systems with tightly packed inner
planets (STIPs) discovered by {\it Kepler}. The scenario involves
build-up of a pebble-dominated protoplanetary ring, supplied by radial
drift of pebbles from the outer disk. It may also involve
further build-up of planetary masses to gap-opening scales via
continued pebble accretion.
Here we study radial drift \& growth of pebbles delivered to the
DZIB in fiducial IOPF disk models.
\end{abstract}

\firstsection % if your document starts with a section,
\section{Introduction}

{\it Kepler} observations show STIPs are very common.
The compact, well-aligned, but largely non-resonant architectures of
these systems challenge formation scenarios involving migration of
already-formed planets from the outer disk (e.g., Hands et al. 2014).

Chiang \& Laughlin (2013) discussed aspects of {\it in situ}
formation. Hansen \& Murray (2012; 2013)
studied STIP formation from a disk of protoplanets,
requiring initial conditions with highly enriched solid surface
densities above the minimum mass solar nebula.

CT14 proposed the IOPF scenario to link enrichment of solids in the
inner disk by pebble drift with simultaneous and sequential formation
of planets at the pressure maximum associated with the
transition from a dead zone to a magneto-rotational instability
(MRI)-dominated zone in the very inner disk, perhaps first set by the
thermal ionization of alkali metals at $\sim 1200$~K.
Here we calculate the radial drift timescale of different sizes of
pebbles starting from various outer disk locations. We then couple
this to a simple growth model for the pebbles. These are the first
steps towards calculation of the global pebble supply rate to the
DZIB, which will control the rate of IOPF.

\section{Drag laws, radial drift velocity and pebble growth}

Following Armitage (2010), 
we define the gas drag frictional timescale of a pebble of mass $m_p$ moving at
speed $v_p$ relative to gas as $t_{\rm fric}=(m_p
v_{p})/F_{D}$. We consider four drag regimes (first is Epstein;
others are Stokes regimes depending on Reynolds no., $\rm Re$):
\begin{equation}
t_{\rm fric}=
\begin{cases}
\rho_p a_p / (\rho_g v_p) & \text{if }  a_p < 9\lambda/4 \\
%\frac{\rho_p a}{\rho_g v} & \text{if }  a < \frac{9}{4}\lambda \\
2\rho_p a_p^2 / (9\nu\rho) & \text{if } a_p > 9\lambda/4 \text{ and } {\rm Re} <1\\
%\frac{2\rho_p a^2}{9\nu\rho} & \text{if } a > \frac{9}{4}\lambda \text{ and } Re <1\\
(\rho_p a_p / [9\rho_g v_{p}]) (2av_{p}/\nu) ^{0.6} & \text{if } 1<{\rm Re}<800\\
%\frac{\rho_p a}{9\rho_g v_{p}}\left(\frac{2av_{p}}{\nu}\right) ^{0.6} & \text{if } 1<Re<800\\
8\rho_p a_p / (1.32\rho_gv_{p}) & \text{if } {\rm Re} > 800,
%\frac{8\rho_pa}{3\rho_gv_{p}}0.44^{-1} & \text{if } Re > 800
\end{cases}
\end{equation}
where
$a_p\equiv a_{p,1}{\rm cm}$ is pebble radius, $\rho_p\equiv\rho_{p,3}3\:{\rm g\:cm^{-3}}$ is pebble density \& $\rho_g$ is gas density.
$\nu \approx \lambda c_s$
is molecular viscosity, 
where $\lambda$ is mean free path of gas molecules \& $c_s$ is sound
speed.  In terms of normalized frictional time, $\tau_{\rm
  fric}\equiv \Omega_K t_{\rm fric}$, where $\Omega_K =
(Gm_*/r^3)^{1/2}$ is orbital angular frequency at radius $r\equiv r_{\rm AU}\:{\rm AU}$
about star of mass $m_*\equiv m_{*,1}M_\odot$, radial drift speed is
\begin{equation}
v_{r,p}  \simeq  -k_P(c_s/v_K)^2 (\tau_{\rm fric}+\tau_{\rm fric}^{-1})^{-1} v_K,
\end{equation}
where $k_P$ describes the disk's pressure profile via
$P=P_0(r/r_0)^{-k_P}$ (with fiducial value of 2.55) and $v_K$ is the
Keplerian velocity.
We 
integrate $v_{r,p}$ to obtain the
radial location of pebbles as a function of time $r_p(t)$.
In the Epstein regime and the Stokes regime with ${\rm Re}<1$, $v_{r,p}$ is solved
for analytically. When ${\rm Re}>1$, we solve numerically. 

We also consider a simple pebble growth model. The pebble accumulates
all small-grain dust material along its path within its physical cross section
while in the Epstein regime. In the Stokes regimes we assume such
accumulation is completely inefficient by deflection of the grains due to the
pressure gradients of diverted gas streamlines.

\section{Results and Discussion}

\begin{figure}[t]
\begin{center}
\includegraphics[width=0.96\textwidth]{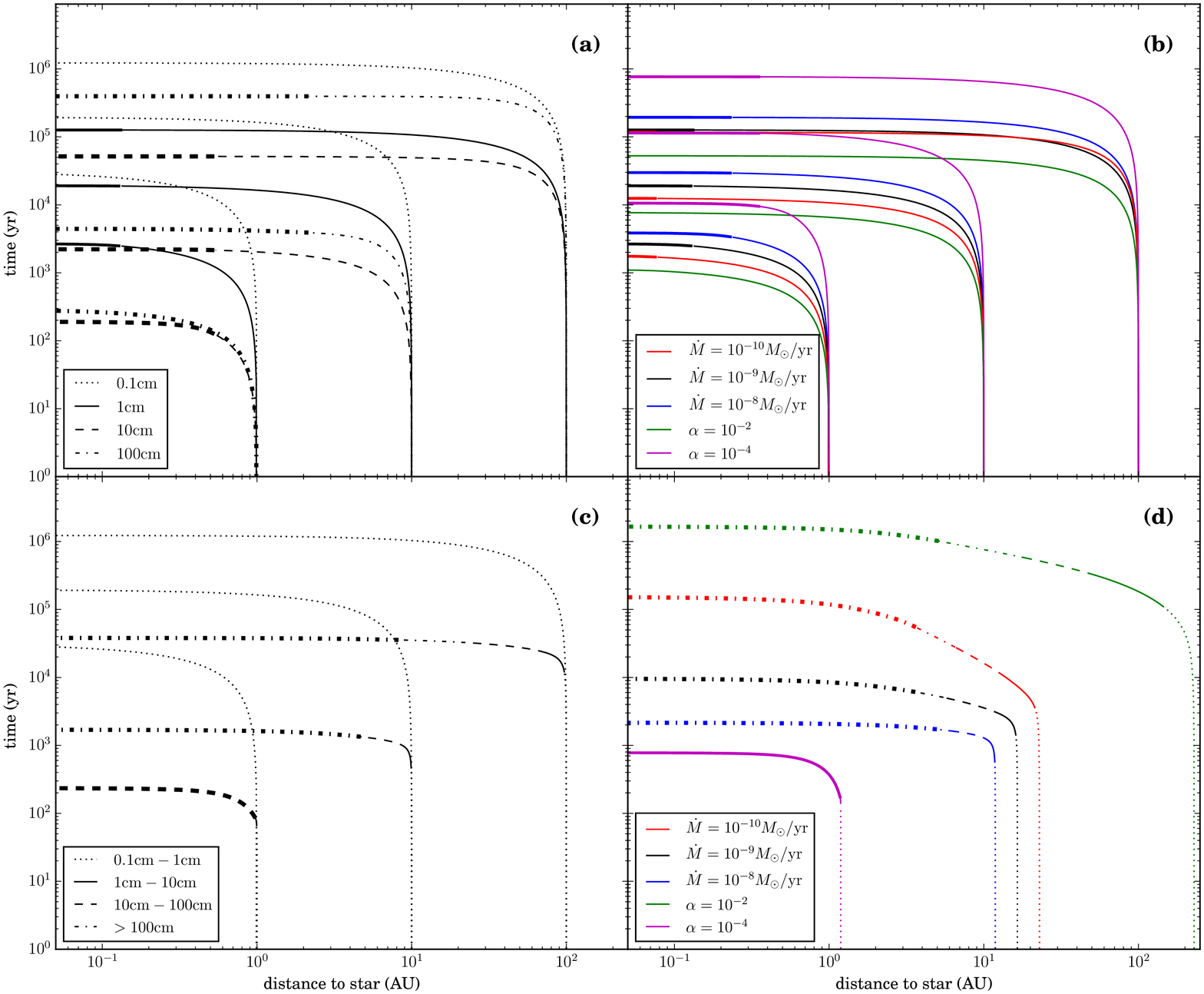}
\caption{
\textbf{a}: Radial drift timescale of pebbles with different fixed
radial sizes starting from various disk radii in the fiducial disk
model ($\dot{m}=10^{-9}\:M_\odot\:{\rm yr}^{-1}$,
$\alpha=10^{-3}$). Thicker lines show the Stokes drag
regimes. \textbf{b}: Radial drift time scale of 1~cm fixed-sized pebbles in disk
models with different accretion rates and $\alpha$ viscosities, varied
around the fiducial model.
\textbf{c}: Comparisons between fixed-size 0.1~cm pebbles and the
growth model with same initial size. The different line styles
indicate growth in pebble radius. \textbf{d}: Growth model drift times
for initially 0.1~cm pebbles starting from the outer radius of the
feeding zone $r_1$ for gap-opening mass planets (see text).
}
\label{fig 1}
\end{center}
\end{figure}

Figure~1a shows examples of radial drift in a fiducial disk model with
accretion rate $\dot{m}\equiv \dot{m}_{-9}10^{-9}\:M_\odot\:{\rm yr}^{-1}$ and
Shakura-Sunyaev viscosity parameter $\alpha\equiv\alpha_{-3} 10^{-3}$
(assumed constant outwards from the DZIB). Pebbles with radius 0.1~cm take
about $10^5$~yr to reach the inner region if starting from 10~AU. This
timescale reduces to about $10^4$~yr and $10^3$~yr for 1~cm and 10~cm
sized pebbles, respectively.

Figure~1b shows that varying the accretion rate over a
range of a factor of 100 around the fiducial value
has only modest effects on the drift timescales. Changing the value of
$\alpha$ over a similar range has a somewhat larger effect. For most
disk radii shown here we are in the Epstein regime with $\tau_{\rm
  fric}\ll 1$. In this case, $v_{r,p}$ is
\begin{equation}
v_{r,p}\simeq 1.92 a_{p,1} \rho_{p,3} \gamma_{1.4}^{8/5}\kappa_{10}^{2/5}\alpha_{-3}^{3/5}m_{*,1}^{-2/5} (f_r\dot{m}_{-9})^{-1/5}r_{\rm{AU}}^{1/5}\:{\rm m\:s^{-1}},
\end{equation}
where $\gamma\equiv1.4\gamma_{1.4}$ is power-law exponent of the
barotropic equation of state, $\kappa\equiv\kappa_{10}10\:{\rm
  cm^2\:g^{-1}}$ is disk opacity, and $f_r\equiv 1-\sqrt{r_*/r}$
(where $r_*$ is stellar radius). So pebbles starting at 10~AU drift
in faster in lower accretion rate disks and in more viscous disks. At
$\sim$100~AU this simple dependence on $\dot{m}$ breaks down as
$\tau_{\rm fric}\gtrsim 1$.

Figure~1c shows the results for the pebble growth model, starting from
0.1~cm sizes. Such pebbles grow quickly, shortening delivery
times from 100~AU to a few $\times 10^4$~yr.

Following CT14, the mass in solids (initially dust) inside disk radius $r_1$ is
\begin{equation}
M_s(<r_1)\approx  0.178 f_{s,-2} \gamma_{1.4}^{-4/5} \kappa_{10}^{-1/5} \alpha_{-3}^{-4/5} m_{*,1}^{1/5} \dot{m}_{-9}^{3/5}  r_{\rm{1,AU}}^{7/5}\:M_\oplus,
\end{equation}
(correcting minor typo in eq. 30 of CT14) where $f_s\equiv
f_{s,-2}0.01$ is the ratio of solid to gas mass. The gap-opening
planetary mass at $r_0$ is
\begin{equation}
M_{G}\approx5.67\phi_{G,0.3}\gamma_{1.4}^{4/5}\kappa_{10}^{-1/5}\alpha_{-3}^{4/5}m_{*,1}^{3/10}(f_r\dot{m}_{-9})^{2/5}r_{\rm{0,AU}}^{1/10}\:M_\oplus,
\end{equation}
based on fraction $\phi_G\equiv\phi_{G,0.3}0.3$ of
viscous-thermal mass (Zhu et al. 2013).  If a
fraction, $\epsilon_p\equiv\epsilon_{p,0.5}0.5$, of disk solid mass
forms pebbles that are accreted by the innermost forming planet with
only minor gas accretion
%and planet formation is dominated by pebble, rather than gas, accretion
(as expected at $T\simeq1200$~K DZIB conditions), then the
required feeding radius of this first gap-opening mass planet is
\begin{equation}
r_1\approx19.4 \phi_{G,0.3}^{5/7}\gamma_{1.4}^{8/7} \kappa_{10}^{2/7} (f_{s,-2}\epsilon_{p,0.5})^{-5/7} \alpha_{-3}^{8/7}m_{*,1}^{1/14}\dot{m}_{-9}^{-1/7}r_{\rm{0,AU}}^{1/14}\:\rm{AU}.
\end{equation}
With fiducial disk parameters and $r_0=0.1$~AU, we have
$r_1=16.5$~AU. The drift time for a 0.1~cm pebble of constant size
from this distance is $2.88\times 10^5$~yr.  The same pebble growing
by sweeping up dust drifts in after $1.15\times 10^4$~yr with final
size of 233~cm. This result is not very sensitive to the choice of
inititial pebble size: a growing pebble starting with 0.01~cm radius
has a drift time of $1.33\times10^4$~yr and a very similar final
size. Note, if $\epsilon_p$ is much smaller than our fiducial value,
e.g., due to interception of pebbles by a population of outer disk
planetesimals (Guillot et al. 2014), then this would increase the
radius of the required feeding zone and thus lengthen the drift
timescales.

These drift times thus set lower limits for the timescale of first,
innermost planet formation in the IOPF model. Figure~1d shows these
times for various $\dot{m}$ \& $\alpha$. They are shorter for disks
that are denser, i.e., due to higher $\dot{m}$ or lower $\alpha$,
given $M_G(\dot{m},\alpha)$ and that the feeding zone is then smaller
and pebble growth more efficient.  These formation times may also be
lower limits if pebble formation by dust grain coagulation (e.g.,
Birnstiel et al. 2012) is the rate limiting step, to be investigated
in future work. Still, Fig.~1d shows IOPF requires $\alpha\lesssim
10^{-3}$, i.e., dead zone conditions, given observed disk lifetimes
$\sim 1$~Myr. Variation in dead zone properties, e.g., from
different disk midplane ionization rates by cosmic rays or
radionuclide decay, could thus lead to a variety of planetary system
formation mechanisms, perhaps helping to explain STIP vs. Solar
System formation.


\begin{thebibliography}{}

\bibitem[Armitage (2010)]{Armitage_07}
{Armitage, P. J.} 2010,
Astrophysics of Planet Formation, Cambridge University Press, 2010
%\textit{arXiv}: 0701485 

\bibitem[Birnstiel,T et.al (2012)]{Birnstiel_2012}
{Birnstiel, T., Klahr, H. \& Ercolano, B.} 2012,
\textit{A\&A}, 539, 148

\bibitem[Chatterjee \& Tan (2014)]{CT14}
{Chatterjee, S., \& Tan, J. C.} 2014, 
\textit{ApJ}, 780, 53

\bibitem[{{Chiang} \& {Laughlin}(2013)}]{2013MNRAS.431.3444C}
{Chiang}, E. \& {Laughlin}, G. 2013, 
\textit{MNRAS}, 431, 3444

\bibitem[Guillot et al ,2014]{Guillot_14}
Guillot, T, Ida, S., \& Ormel, C. W. 2014,
\textit{arXiv}: 1409.7328

\bibitem[Hands,T.O. et al ,2014]{Hands_14}
{Hands, T. O., Alexander, R. D. \& Dehnen, W.} 2014,
\textit{arXiv}: 1409.0532

\bibitem[{{Hansen} \& {Murray}(2012)}]{2012ApJ...751..158H}
{Hansen}, B.~M.~S. \& {Murray}, N. 2012,
\textit{ApJ}, 751, 158

\bibitem[{{Hansen} \& {Murray}(2013)}]{2013arXiv1301.7431H}
{Hansen}, B.~M.~S. \& {Murray}, N. 2013, 
\textit{ApJ}, 775, 53

\bibitem[{Zhu}]{}
Zhu, Z., Stone, J. M., \& Rafikov, R. R. 2013
\textit{ApJ}, 768, 143


\end{thebibliography}
\end{document}